\documentclass[12pt,a4paper,english]{article}

\usepackage{float}
\usepackage[slantedGreek]{mathpazo}
\usepackage[pdftex]{graphicx}
\usepackage{amsfonts}
\usepackage{setspace}
\usepackage{amssymb}
\usepackage{amsthm}
\usepackage{graphicx}
\usepackage{amsmath,scalerel}%
\usepackage{subcaption}
\usepackage{color}
\usepackage[english]{babel}
 
\usepackage{natbib}
\usepackage{graphicx}
\usepackage{amsmath, amsthm, amssymb, amsfonts}
\usepackage{bm}
\usepackage{epstopdf}
\usepackage{times}
\usepackage{hyperref}
\usepackage{booktabs}
\usepackage{array}
\usepackage{lscape}
\usepackage{tikz}
\usepackage{algorithm2e}
\usepackage{caption}
\newtheorem{lemma}{Lemma}[section]
\usepackage[all]{hypcap}                    
\usepackage{babel}

\DeclareMathOperator*{\argmax}{arg\,max}
\DeclareMathOperator*{\diag}{diag}
\DeclareMathOperator{\Var}{Var}

\newcommand{\matr}[1]{\bm{#1}} 

\usepackage{epstopdf}

\theoremstyle{plain}
\newtheorem{definition}{Definition}[section]

\title{Dynamic sparsity on time-varying parameter models}
\author{Paloma Vaissman Uribe\\
Hedibert Freitas Lopes}

\begin{document}

\begin{center}

{\Large\sc Dynamic sparsity on dynamic}

\vspace{0.2cm}

{\Large\sc regression models}


\vspace{1.5cm}

\singlespacing

{\large Paloma Vaissman Uribe}\\
Data Scientist, iFood, S\~ao Paulo, Brazil

\vspace{0.4cm}
and
\vspace{0.4cm}

{\large Hedibert Freitas Lopes}\\
Insper, S\~ao Paulo, Brazil
\end{center}

\vspace{1.5cm}

\noindent In the present work, we consider variable selection and shrinkage for the Gaussian dynamic linear regression within a Bayesian framework. In particular, we propose a novel method that allows for time-varying sparsity, based on an extension of spike-and-slab priors for dynamic models. This is done by assigning appropriate Markov switching priors for the time-varying coefficients' variances, extending the previous work of \cite{ishwaran2005spike}. Furthermore, we investigate different priors, including the common Inverted gamma prior for the process variances, and other mixture prior distributions such as Gamma priors for both the spike and the slab, which leads to a mixture of Normal-Gammas priors (\cite{griffin2010inference}) for the coefficients. In this sense, our prior can be view as a dynamic variable selection prior which induces either smoothness (through the slab) or shrinkage towards zero (through the spike) at each time point. The MCMC method used for posterior computation uses Markov latent variables that can assume binary regimes at each time point to generate the coefficients' variances. In that way, our model is a dynamic mixture model, thus, we could use the algorithm of \cite{gerlach2000efficient} to generate the latent processes without conditioning on the states. Finally, our approach is exemplified through simulated examples and a real data application.\\




\section{Introduction}

Over the past few decades, advances in computational processing have encouraged the proliferation of massive datasets, bringing new challenges to statistical research due to high-dimensionality issue. In this sense, regularization and variable selection techniques have become even more relevant to prevent overfitting and to solve ill-posed problems by inducing sparsity and/or shrinkage. 

The advantage of regularization was highlighted a few years ago when \cite{elements} coined the informal \textit{Bet on Sparsity} principle. The principle encourages the use of procedures that do well in sparse problems for high-dimensional problems, since no procedure does well in dense problems, accordingly to them. Indeed, they have shown that for a dense problem, where all the numerous coefficient where different from zero, and/or there is a high \textit{Noise-to-Signal Ratio} (NSR), both the former \textit{ridge regression} procedure of \cite{hoerl1970ridge} and the \textit{least absolute shrinkage and selection operator} (lasso) from \cite{tibshirani1996} do poorly in terms of prediction. 


Consider the Gaussian linear model defined by

\begin{equation}\label{linreg1}
\matr{y}=\beta_0\matr{1}+\matr{X}\matr{\beta}+\matr{\varepsilon}, \qquad \matr{\varepsilon} \sim  \mathcal{N}(\matr{0},\sigma^2\matr{I}),
\end{equation}

\noindent where $\matr{y}$ is a $n$-dimensional vector of continuous responses, $\beta_0$ is the intercept, $\matr{\beta}$ is a $q$-dimensional vector of regression coefficients associated with covariates, $\matr{X}$ is a $(n \times q)$ design matrix with each column representing a covariate.

It is well known that the \textit{ordinary least squares} (OLS) or \textit{maximum likelihood estimator} (MLE) $\hat{\beta}_{OLS}=(\matr{X}'\matr{X})^{-1}\matr{X}'\matr{y}$ often does poorly on both interpretation and prediction accuracy. While interpretation stands for the preference for parsimony, in the sense that simpler models put more light on the relationship between the response and covariates, prediction accuracy is related to the bias-variance trade-off. Although the OLS/ MLE estimator has the smallest variance among all linear unbiased estimators accordingly to the Gauss Markov Theorem, an estimator with slight bias but smaller variance could be preferable, leading to a substantial decrease in prediction error. 

Modern statistics addresses this trade-off between bias and variance through regularization and variable selection methods, which encourages simpler models because the space of values of estimators $\hat{\matr{\beta}}$ considered is smaller. In general terms, the notion of regularization summarizes approaches that allow to solve ill-posed problems, such as those which arises when $p \gg n$, or to prevent overfitting. The problem of variable selection arises when there is some unknown subset of the predictors with regression coefficients so small that it would be preferable to ignore them. 

Recently, there has been great interest in regularizing the coefficients within problems in which the parameters vary over time and a few methods were proposed such as those from \cite{belmonte2014hierarchical},  \cite{kalli2014time} and \cite{bitto2016achieving}. In linear regression models with a large number of predictors, it is common to assume that only a subset of them is important for prediction. In the context of dynamic regression, it is reasonable to assume that these relevant subsets change over time. 

Actually, we can define two sources of sparsity in dynamic regression problems: the \textit{vertical sparsity} or \textit{dynamic sparsity}, which stands for time-varying subsets of relevant predictors, and the \textit{horizontal sparsity}, which allows for intermittent zeros for when each individual predictor is not relevant at all times $t$. 

Consider the Gaussian dynamic linear regression model defined by

\begin{equation}\label{dynreg}
\begin{aligned}
y_t&=\matr{F}'_t\matr{\beta}_t+\nu_t, \qquad \nu_t \sim \mathcal{N}(0,\sigma_t^2),\\
\matr{\beta}_t&=\matr{G}_t\matr{\beta}_{t-1}+\matr{\omega}_t, \qquad \matr{\omega}_t \sim \mathcal{N}(\matr{0},\matr{W}_t),
\end{aligned}
\end{equation}

\noindent for $t=1,\ldots,T$, where $\matr{F}_t'=\matr{X}_t$ is the $(1 \times q)$ vector of regressors, $\matr{\beta}_t=(\beta_{1t},\ldots,\beta_{qt})$ is the $(q \times 1)$ vector of coefficients and $\nu_t$ and $\matr{\omega}_t$ are two independent sequences of independent Gaussian errors with mean zero and variances $\sigma_t^2$ and $\matr{W}_t$, respectively. Note that setting $\matr{W}_t=\matr{0}$ for all $t$ is equivalent to $\matr{\beta}_t=\matr{\beta}$, i.e., the static regression. Usually, we may define $\matr{G}_t=\matr{G}=\matr{I}_q$.  

The problem of shrinking in dynamic regression problems has been addressed by \cite{belmonte2014hierarchical} and \cite{bitto2016achieving} in a similar strategy.  See also \cite{huber2020} for recent comparative study.  Their basic approach was to rewrite the states $\beta_{j,t}$ from \eqref{dynreg} in terms of the scaled states as $\tilde{\beta}_{j,t}=\beta_{j,t}/\omega_j$. Then, the shrinkage of the time-varying coefficients $\beta_{j,t}$ was done by assigning priors to their standard deviations $\omega_j$. While \cite{belmonte2014hierarchical} used the Laplace prior for shrinking the standard deviations, \cite{bitto2016achieving} used the Normal-Gamma prior, which is more general since the Laplace prior is a special case of the Normal-Gamma prior where the shrinking parameter equals one. 

In both approaches, the standard deviation $\omega_j$ plays the role of relevance of the $j$th predictor: small values of $\omega_j$ leads to greater shrinkage of the coefficient $\beta_{j,t}$ for all times $t$. That is, because the standard deviation $\omega_j$ is taken as fixed for all times $t$, if it is pulled toward zero, then the time-varying effect of the covariate $\matr{X}_j$ is non significant. In this sense, both tackles horizontal sparsity, as the shrinkage effect of the prior for $\omega_j$ is equal over all times $t$.   

On the other hand, \cite{kalli2014time}, developed an extension of the Normal-Gamma prior discussed by \cite{griffin2010inference} for dynamic regression where both the values of the regression coefficients and the importance of the variables are allowed to change over time. Time-varying sparsity is allowed by giving independent Normal-Gamma autoregressive (NGAR) process priors to the time series of regression coefficients. For details on the model specification, please refer to \cite{kalli2014time}.  See also \cite{lopes2008}, who propose a customized four-component mixture prior for the elements of $\matr{\beta}_t$.  The prior induces sparsity by either flatlining the coefficient to a constant, possibly zero, or letting move around freely according the dynamic structure when the state vector is of very high dimensions, say five thousand equations.  All these methods deal with horizontal sparsities.

Vertical sparsity in dynamic models has become an increasingly interesting research topic.   Related literature include
 \cite{nakajima2013bayesian}, \cite{kalli2014time}, \cite{rockovamcalinn2018} and \cite{kowal2018} and, more recently,  
 \cite{koop2020} who uses variational Bayes ideas.

%
%

In the present work, we consider variable selection and shrinkage for the Gaussian dynamic linear regression model within a Bayesian framework. In particular, we propose a method that allows for time-varying sparsity, based on an extension of spike-and-slab priors for dynamic models using Markov switching auxiliary variables. The paper is organized as follows: Section \ref{shrinkage} reviews the Bayesian approach for regularization and variable selection, emphasizing the main shrinking priors and giving a unifying approach for the spike-and-slab priors. Section \ref{model} introduces the dynamic spike-and-slab prior and describes the full Bayesian model for dynamic regression with time-varying sparsity, the posterior inference and the Markov chain Monte Carlo (MCMC) method for sampling the parameters. Section \ref{simulations} applies the proposed model to simulated data and considers empirical studies in inflation modeling. Section \ref{conclusion} summarizes our findings and conclusions.

\section{Shrinking and variable selection priors: a brief review}\label{shrinkage}

Classical regularization is done by maximizing the likelihood subject to a penalty function. While the the ridge regression estimate is a penalized least squares method imposing a $\ell_2$ penalty on the regression coefficients, the lasso uses the $\ell_1$ norm instead, leading not only to shrunken coefficients but also to sparse solutions. In that way, the lasso is also considered a variable selection method. 

In contrast, Bayesian approaches for both variable selection and regularization in the Gaussian linear model stated by \eqref{linreg1} can be formalized through the conditional distribution $p(\matr{y}|\matr{\beta},\matr{\phi})$, where $\matr{\phi}$ is a parameter vector, comprising, for example, the error variance $\sigma^2$. The regularization is achieved by specifying appropriate informative priors $p(\matr{\beta}|\matr{\theta})$, where the hyperparameter vector $\matr{\theta}$ includes parameters controlling shrinkage properties. The model is completed by assuming hyperpriors $p(\matr{\theta})$ and $p(\matr{\phi})$ and the inference is based on the posterior $p(\matr{\beta},\matr{\phi},\matr{\theta},\matr{y}) \propto p(\matr{y}|\matr{\beta},\matr{\phi})p(\matr{\beta}|\matr{\theta})p(\theta)p(\matr{\phi})$.

It can be shown that, if $\matr{\phi}$ and $\matr{\theta}$ are fixed, then the posterior mode or the maximum a posteriori (MAP)
$$\underset{\beta}\argmax\{p(\matr{y}|\matr{\beta},\matr{\phi})p(\matr{\beta}|\matr{\theta})\}$$

\noindent is equivalent to penalizing the log-likelihood $\log{p(\matr{y}|\matr{\beta},\matr{\phi})}$ with penalty equal to the (minus) log prior $\log p(\matr{\beta}|\matr{\theta})$. 

\subsection{Shrinking priors}

In general, practically all shrinking priors are defined hierarchically as a \textit{Scale-Mixture of Normals} (SMN) (see, e.g., \cite{west1987scale}). Considering the Gaussian linear model as in \eqref{linreg1}, the SMN has the following general structure:

\begin{equation}\label{smn1}
\beta_j|\psi_j \stackrel{\text{ind}} \sim \mathcal{N}(0,\psi_j), \qquad \psi_j|\theta \sim p(\psi_j|\matr{\theta}),
\end{equation}

\noindent where $\beta_i$ and $\beta_j$ are independent for any $i, j \in \{1,\ldots,q\}$ and $\psi_j$ depends on the vector of hyperparameters $\matr{\theta}$. Note that the marginal distribution $p(\beta_j|\matr{\theta})=\int p(\beta_j|\psi_j)p(\psi_j|\matr{\theta})d\psi_j$ is non Gaussian, and can assume many forms depending on the mixing distribution $p(\psi_j|\matr{\theta})$. A famous form arises when the mixing distribution $p(\psi_j|\matr{\theta})$ from \eqref{smn1} is Exponential, that is,

\begin{equation*}\label{bayeslasso2}
\psi_j|\lambda \sim \mathcal{E}\left(\lambda^2/2\right),
\end{equation*}

\noindent for $j=1,\ldots,q$, where $\mathcal{E}(\alpha)$ denotes the Exponential distribution with mean $1/\alpha$. Thus, marginally, $\beta_j$ follows a Laplace distribution with parameter $\lambda$, so that $p(\beta_j) \propto \exp(-\lambda|\beta_j|)$. The Laplace prior is also known as the Bayesian Lasso from \cite{park2008bayesian}.

The frequentist ridge regression also has a Bayesian analogue and can be represented by \eqref{smn1} if we consider an Inverted-Gamma mixing distribution $p(\psi_j|\matr{\theta})$. That is, assuming
\begin{equation*}\label{bayesridge}
\psi_j|a,b \sim  \mathcal{IG}(a,b),
\end{equation*}

\noindent for $j=1,\ldots,q$, then, $\beta_j$ follows a scaled $t$ distribution with $2a$ degrees of freedom and scale parameter $\sqrt{a/b}$, marginally. Hence, the ridge prior leads to weaker penalization of large coefficients as long as the $t$ distribution has heavier tails than the Gaussian distribution. 

\subsection{The Normal-Gamma prior}

Even more shrinkage than the Bayesian lasso and the ridge prior can be achieved by using a Gamma mixing distribution in \eqref{smn1} as
 
\begin{equation*}\label{normalgamma1}
\psi_j|\lambda, \gamma^2 \sim  \mathcal{G}(\lambda,1/(2\gamma^2)),
\end{equation*}

\noindent for $j=1,\ldots,q$, where $\mathcal{G}(\lambda,1/(2\gamma^2))$ denotes the Gamma distribution with shape parameter $\lambda$ and mean $2\lambda \gamma^2$. This structure leads to the Normal-Gamma density, which was applied to regression problems in \cite{griffin2010inference} and can be expressed in closed form as

\begin{equation}\label{normalgamma2}
p(\beta_j)=\dfrac{1}{\sqrt{\pi} 2^{\lambda-1/2}\gamma^{\lambda+1/2} \Gamma(\lambda)} |\beta_j|^{\lambda-1/2}K_{\lambda-1/2}(|\beta_j|/\gamma),
\end{equation} 

\noindent where $K$ is the modified Bessel function of the third kind. Taking this parametrization, the variance of $\beta_j$ is $2\lambda \gamma^2$ and the excess kurtosis is $3/\lambda$. As the shape parameter $\lambda$ of the Gamma distribution decreases these include distributions that place a lot of mass close to zero but at the same time have heavy tails. Thus, the effect of the parameter $\lambda$ is related to shrinking, where lower values of $\lambda$ is associated with more shrinkage as more mass is placed to zero. Thusly, the Normal-Gamma prior is more general than the Bayesian lasso, which corresponds to $\lambda=1$.

One could fix the shrinking parameter $\lambda$ such as in \cite{fruhwirth2010stochastic} and in \cite{kastner2016sparse}\footnote{In \cite{fruhwirth2010stochastic}, they set $\lambda=1/2$, in which case $\psi_j|\gamma^2 \sim \gamma^2\chi_1^2$, or equivalently, $\sqrt{\psi_j} \sim N(0,\gamma^2)$. In \cite{kastner2016sparse}, it was assumed that $\lambda=0.1$.}
 or adopt a fully Bayesian approach by assigning hyperpriors to both $\lambda$ and $\gamma^2$ as did \cite{griffin2010inference}. A prior which seemed to work well in the simulations is taking $\lambda$ to be an exponential distribution with mean 1, which offers variability around the Bayesian lasso.
 
\subsection{Variable selection: spike-and-slab priors}\label{ssp}

Bayesian variable selection is commonly based on spike-and-slab priors for regression coefficients. The basic idea is that each component $\beta_j$ from the coefficients' vector $\matr{\beta}$ is modeled as having come either from a distribution with most (or all) of its mass concentrated around zero (the \textit{spike}), or from a comparably diffuse distribution with mass spread out over a large range of values (the \textit{slab}), that is
\begin{equation}\label{generic_ss}
\beta_{j}|J_{j} \sim J_{j} p_{slab}(\beta_{j}|\matr{\theta})+(1-J_{j})p_{spike}(\beta_{j}|\matr{\theta}),
\end{equation}

\noindent where $\matr{\theta}$ is a vector of parameters, $p_{spike}(\beta_{j}|\matr{\theta})$ is the spike distribution, $p_{slab}(\beta_{j}|\matr{\theta})$ is the slab distribution and $J_j \in\{0,1\}$ is a binary random variable with $p(J_j=1)=1-P(J_j=0)=\omega$.

Famous seminal \textit{stochastic search variable selection} (SSVS) method of \cite{george1993variable} used Gaussian distributions for both the spike and the slab as follows
\begin{equation}\label{ssvs1a}
\beta_j|J_j  \sim (1-J_j)\mathcal{N}(0,\tau_j^2)+J_j\mathcal{N}(0,c_j^2\tau_j^2), 
\end{equation}

\noindent for $j=1,..,q$, where $c_j >1$ is a large scalar and $\tau_j >0$ is a small scalar. Note that $c_j$ is the ratio of variances between the slab and the spike distributions.

Indeed, one can achieve \eqref{generic_ss} by formulating appropriate spike-and-slab priors to the component variances $(\psi_j|J_j=0)=\Var_{spike}(\beta_{j}|\matr{\theta})$ and $(\psi_j|J_j=1)=\Var_{slab}(\beta_{j}|\matr{\theta})$, what was first proposed by \cite{ishwaran2005spike}. Their idea is that, considering the hierarchical SMN representation from \eqref{smn1}, if we assume absolutely continuous priors for the component variances, that is, a mixture prior for $\psi_j$, then we reach the spike-and-slab structure for the coefficient $\beta_j$ as in \eqref{generic_ss}. 

Actually, even the SSVS approach given by \eqref{ssvs1a}, where the coefficients are given a mixture of Normals prior, can be represented in that way. Letting $c_j=1/r=\Var_{slab}(\beta_{j}|\matr{\theta})/\Var_{spike}(\beta_{j}|\matr{\theta})$ and $\tau_j^2=r$ from \eqref{ssvs1a} and assuming a two point mixture prior to the variance $\psi_j$ in \eqref{smn1}
$$\psi_j|\omega,Q,r \sim (1-\omega) \delta_{rQ}(.)+\omega\delta_{Q}(.),$$

\noindent where $\delta_{\upsilon}(.)$ is a discrete measure concentrated at value $\upsilon$, we get the original formulation of the SSVS prior of \cite{george1993variable}.

Nevertheless, it can be difficult to set the hyperparameters $r$, $Q$ and $\omega$ parameters used in the two point mixture prior for the variance $\psi_j$, as noted by \cite{ishwaran2005spike}. Hence, their approach is to place absolutely continuous priors for $\psi_j$. In particular, they chose Inverted-Gamma densities for both the spike and the slab variances as
$$\psi_j|J_j=0 \sim \mathcal{IG}(\nu,rQ), \qquad \psi_j|J_j=1 \sim\mathcal{IG}(\nu,Q),$$

\noindent where $\matr{\theta}=(\nu,r,Q)$ is the vector of hyperparameters that define the conditional (on $J_j$) variances' densities. Thus, each $\beta_j$ has the marginal distribution 
\begin{equation}\label{nmig}
\beta_j|\omega,r,Q \sim \omega t_{2 \nu}(0,Q/\nu)+(1-\omega)t_{2 \nu}(0,rQ/\nu),
\end{equation}

\noindent where $t_{\xi}(0,s)$ denotes the Student's t distribution with zero location, scale $\sqrt{s}$ and $\xi$ degrees of freedom. Because of the SMN representation, \eqref{nmig} is also know as \textit{Normal mixture of Inverse-Gamma} (NMIG) prior. Although it allows discrimination or variable selection, it does not encourage shrinkage in the sense that the resulting marginal distribution of each coefficient $\beta_j$ is a two component mixture of scaled Student's t distributions. Hence it makes sense to choose other component specific distributions, besides the Inverse Gamma, that could actually induce shrinkage. A prior that solves this is assuming Exponential densities for both the spike and the slab as 
$$\psi_{j}|J_{j}=0 \sim \mathcal{E}(1/2rQ), \qquad \psi_{j}|J_{j}=1 \sim \mathcal{E}(1/2Q),$$

\noindent where $\mathcal{E}(\alpha)$ denotes the Exponential distribution with mean $1/\alpha$, which leads to 

\begin{equation}\label{laplace2}
\beta_{j}|\omega,r,Q \sim  \omega Lap(\sqrt{Q})+(1-\omega)Lap(\sqrt{rQ}), 
\end{equation}

\noindent where $Lap(x)$ denotes the Laplace distribution with mean 0 and scale parameter $x$ and the weight $\omega$ is the prior probability of the slab, i.e., $\omega=p(J_{j}=1)$. That is, \eqref{laplace2} is a mixture of Laplace densities for $\beta_{j}$. Finally, if we assume that $\psi_j$ is a mixture of Gammas, that is,
$$\psi_{j}|J_j=0 \sim \mathcal{G}(a,1/2rQ), \qquad \psi_{j}|J_j=0 \sim \mathcal{G}(a,1/2Q),$$

\noindent then the marginal distribution of $\beta_{j}$ is a mixture of Normal-Gamma densities as showed in \eqref{normalgamma2}, that is,

\begin{equation}\label{NG2}
\beta_{j}|\omega,r,Q \sim  \omega \mathcal{NG}(\beta_{j}|a,Q)+(1- \omega)\mathcal{NG}(\beta_{j}|a,r,Q). 
\end{equation}

Hence, assuming either \eqref{nmig}, \eqref{laplace2} or \eqref{NG2}, we adopt a unifying approach for spike-and-slab priors as dicussed in \cite{bernardo2011bayesian}. Taking this parametrization, we will always have
$$\Var_{spike}(\beta_j|r,Q)=cQr, \qquad \Var_{slab}(\beta_j|Q)=cQ,$$

\noindent with $c$ being a constant which depends on the distribution assumption. Table \ref{tab:tab_mixture_priors} gives a summary for what as discussed through this section assuming the general form from \eqref{generic_ss} and viewing each prior as a scaled mixture of Normals (SMN). 

\vspace{0.4cm}

\begin{table}[ht]
\centering
\resizebox{1 \textwidth}{!}{
\begin{tabular}{@{}lccccc@{}}
\toprule
\textbf{Prior }                   & \textbf{Spike $\psi|J=0$} & \textbf{Slab $\psi|J=1$ }& \textbf{Marginal $\beta|\omega$} & \textbf{Constant $c$} \\ \midrule
SSVS                  &   $ \psi|J=0 = \delta_{rQ}(.)  $         &   $\psi|J=1 = \delta_{Q}(.) $           &    $\omega \mathcal{N}(0,Q)+(1-\omega) \mathcal{N}(0,rQ)$        &  1 \\ 
NMIG                  & $ \mathcal{IG}(\nu,rQ)$  & $\mathcal{IG}(\nu,Q)$ &   $\omega t_{2\nu}(0,Q/\nu)+(1-\omega) t_{2\nu}(0,rQ/\nu)$ & $1/(\nu-1$)  \\ 
Mixture of Laplaces      &    $\mathcal{E}(1/2rQ) $           &  $\mathcal{E}(1/2Q)$        &   $\omega Lap(\sqrt{Q})+(1-\omega)Lap(\sqrt{rQ}) $        & 2   \\ 
Mixture of Normal-Gammas &     $\mathcal{G}(a,1/2rQ) $          &    $\mathcal{G}(a,1/2Q)$      &    $\omega \mathcal{NG}(\beta_{j}|a,Q)+(1- \omega)\mathcal{NG}(\beta_{j}|a,r,Q)$   & $2a$   \\ 
Laplace-t                &   $\mathcal{E}(1/2rQ)$              &  $\mathcal{IG}(\nu,Q) $       &   $\omega t_{2\nu}(0,Q/\nu)+(1-\omega)Lap(\sqrt{rQ})$        &  $c_1=2, \quad c_2=1/(\nu-1)$ \\\bottomrule
\end{tabular}}
\vspace{0.4cm} 
\caption{Summary table of an unifying approach for spike-and-slab mixture priors. Depending on the assumption for the mixture prior of $\psi_j$, the constant $c$ changes so that we can compare the different priors, fixing the value of the component variances $\Var_{spike}(\beta_j|r,Q)$ and $\Var_{slab}(\beta_j|Q)$.}
\label{tab:tab_mixture_priors}
\end{table}

In this Section we have discussed several shrinkage and sparsity inducing priors independently assigned to static coefficients from the Gaussian linear model. Now we turn attention to the case where the coefficients from regression are time-varying, the so-called time-varying parameter (TVP) regression models. Next section describes the proposed model for shrinking and selecting subsets of relevant variables dynamically. 

\section{Our approach for sparsity in TVP models}\label{model}

The aim of the method is allowing for vertical sparsity, so that in each snapshot of time $t$ we can have different subsets of relevant predictors as well enabling shrinkage of the time-varying coefficients. This is accomplished by extending the approach of spike-and-slab priors over the variances discussed in Section \ref{ssp} for dynamic models. In our approach, at each time point, an auxiliary Markov switching variable can assume two regimes -  the spike or the slab - driving the evolution of the variance $\psi_{j,t}$, assuming that $\beta_{j,t}|\psi_{j,t} \sim \mathcal{N}(0,\psi_{j,t})$.

\subsection{Model specification} 
\label{subsec:dynamic_spike_model}

The observation and the state equations considered are given below. We work with univariate time series responses, although extending the framework to multivariate time series is straightforward. We assume the Gaussian dynamic linear regression model

\begin{equation}\label{dyn_nmig0}
y_t = \matr{X}_t\matr{\beta}_t+\nu_t, \qquad \nu_t \sim \mathcal{N}(0,\sigma_t^2),
\end{equation}

\noindent for $t=1,\ldots,T$, where $\matr{X}_t$ is a $(T \times q)$ matrix of regressors, $\matr{\beta}_t$ is a $(q \times 1)$ vector of coefficients with the following evolution equation for the scaled states $\tilde{\matr{\beta}}_{1:T}=(\tilde{\matr{\beta}}_1,\ldots,\tilde{\matr{\beta}}_T)$. 

\begin{equation}\label{dyn_nmig1}
\tilde{\matr{\beta}}_t = \matr{G}_t\tilde{\matr{\beta}}_{t-1}+\matr{\eta}_t, \qquad \matr{\eta}_t \sim \mathcal{N}(\matr{0},\matr{W}_t),
\end{equation}

\noindent for $t=2,\ldots,T$, with

\begin{equation*}
\begin{aligned}
&\tilde{\matr{\beta}}_t=(\beta_{1,t}/\sqrt{\psi_{1,t}},\ldots,\beta_{q,t}/\sqrt{\psi_{q,t}}),\\
&\matr{G}_t=\diag(\phi_{1},\ldots,\phi_q),\\
&\matr{W}_t=\diag((1-\phi_1^2),\ldots,(1-\phi_q^2)),
\end{aligned}
\end{equation*}

\noindent where the initial condition for the scaled states is $\tilde{\matr{\beta}}_1 \sim \mathcal{N}(\matr{0},\matr{I})$.

The definition below specifies the generic dynamic spike-and-slab prior that can be assigned to the coefficients' variances $\matr{\psi}_{j,1:T}$ in order to induce shrinkage and/or variable selection.

\begin{definition}{\textbf{Dynamic spike-and-slab prior.}}\label{dyn_spike_def} Consider that $\psi_{j,t}=K_{j,t}\tau_j^2$. The dynamic spike-and-slab prior for $\beta_{j,1:T}$ is defined by  \eqref{dyn_nmig0}, \eqref{dyn_nmig1} and 

\begin{equation}\label{dyn_nmig2}
\begin{aligned}
\tau_j^2 &\stackrel{\text{iid}}\sim p(\tau_j^2|\matr{\theta}),\\
(K_{j,t}|K_{j,t-1}=\upsilon_i) &\stackrel{\text{ind}} \sim \omega_{j,1,i}\delta_{1}(.)+(1-\omega_{j,1,i})\delta_{r}(.),\\
\omega_{j,1,i} &= p\left(K_{j,t}=1|K_{j,t-1}=\upsilon_i\right),
\end{aligned}
\end{equation}

\noindent for $j=1,\ldots,q$, $t=2,\ldots,T$, where $\delta_x(.)$ is a discrete measure concentrated at value $x$, $\upsilon_i \in \{r,1\}$, $p(K_{j,1}=r)=p(K_{j,1}=1)=1/2$ and $p(\tau_j^2|\matr{\theta})$ can be one of the mixing distributions from Table ~\ref{tab:tab_mixture_priors}. As in Section \ref{ssp}, we assume that $r=\Var_{spike}(\beta_{j}|\matr{\theta})/\Var_{slab}(\beta_{j}|\matr{\theta}) \ll 1$ and that $\upsilon_1=1$. 

\end{definition}

Because now we are talking about dynamic models and dynamic sparsity, we have a time-varying scale parameter $\psi_{j,t}$, which is taken to be $\psi_{j,t}=K_{j,t}\tau_j^2$. That is, the time-varying pattern for the scale parameter is driven by the latent variable $K_{j,t}$, which evolves as a Markov switching process of order 1 and can assume two values $K_{j,t}=1$ or $K_{j,t}=r$ accordingly to a transition matrix. Thus, we assume a finite mixture prior for $\psi_{j,t}$ as
\begin{equation*}\label{dyn_nmig4}
\psi_{j,t}|K_{j,t-1}=\upsilon_i \sim  \omega_{j,1,i}p_{slab}(\psi_j|Q_j)+(1- \omega_{j,1,i})p_{spike}(\psi_j|r,Q_j),
\end{equation*} 
 
\noindent where $\omega_{j,1,i}$ is the transition probability of the first order Markov process $K_{j,t}$ to regime $K_{j,t}=1$ (the slab) given that $K_{j,t-1}=\upsilon_i \in \{r,1\}$. Thus, by adopting a regime switching model, the process $\psi_{j,t}$ can switch between the spike and the slab variances' distributions according to the following transition probabilities 
$$
   \matr{\mathcal{P}}_j=
  \left[ {\begin{array}{cc}
   \omega_{j,0,0} & \omega_{j,0,1} \\
   \omega_{j,1,0} & \omega_{j,1,1} \\
  \end{array} } \right]
  $$

\vspace{0.2cm}

\noindent where $\omega_{j,k,i}=P(K_{j,t}=\upsilon_k|K_{j,t-1}=\upsilon_i)$ denotes the probability of $K_{j,t}$ changing to regime $\upsilon_k$ from regime $\upsilon_i$, $k,i \in \{0,1\}$. Note that $\omega_{j,0,1}=(1-\omega_{j,1,1})$ and $\omega_{j,1,0}=(1-\omega_{j,0,0})$.

For the other component $\tau_j^2$ is placed a prior distribution (Inverse-Gamma, Gamma or Exponential) that together with the variable $K_{j,t}$ results in a spike-and-slab prior for $\psi_{j,t}$ that shrinks the coefficients $\beta_{j,t}$ whenever it gets a small value through the spike component of the mixture prior. 

For instance, if $p(\tau_j^2|\matr{\theta}) \sim \mathcal{IG}(\nu,Q_j)$, then we have a mixture of scaled-t for each $\beta_{j,t}$ as
\begin{equation}\label{marg_beta_NMIG}
\beta_{j,t}|\omega_{j,1,i} \sim \omega_{j,1,i} t_{2 \nu}(0,Q_j/\nu)+(1-\omega_{j,1,i})t_{2\nu}(0,rQ_j/\nu).
\end{equation}

If $p(\tau_j^2|\matr{\theta}) \sim \mathcal{G}(a_\tau,1/2Q_j)$, then we have a mixture of Normal-Gamma densities for each $\beta_{j,t}$ as
\begin{equation}\label{marg_beta_NG}
\beta_{j,t}|\omega_{j,1,i} \sim \omega_{j,1,i} \mathcal{NG}(\beta_{j,t}|a_\tau,Q_j)+(1-\omega_{j,1,i})(\beta_{j,t}|a_\tau,r,Q_j).
\end{equation}

Finally, if $p(\tau_j^2|\matr{\theta}) \sim \mathcal{E}(1/2Q_j)$, then we have a mixture of Laplaces densities for each $\beta_{j,t}$ as
\begin{equation}\label{marg_beta_Laplace}
\beta_{j,t}|\omega_{j,1,i} \sim \omega_{j,1,i} Lap(\sqrt{Q_j})+(1-\omega_{j,1,i})Lap(\sqrt{rQ_j}).
\end{equation}

Furthermore, we assume that $\psi_{j,t} \sim \mathcal{IG}(c_{\psi},C_{\psi})$, which means that $Q_j$ is distributed as 
\begin{equation}\label{priorvar_dyn}
Q_j|\omega_{j,1,i} \stackrel{\text{ind}} \sim \mathcal{IG}(c_\psi,C_\psi/f^{*}(w)),
\end{equation}

\begin{figure}[ht!]
\centering
\includegraphics[width=0.75\textwidth]{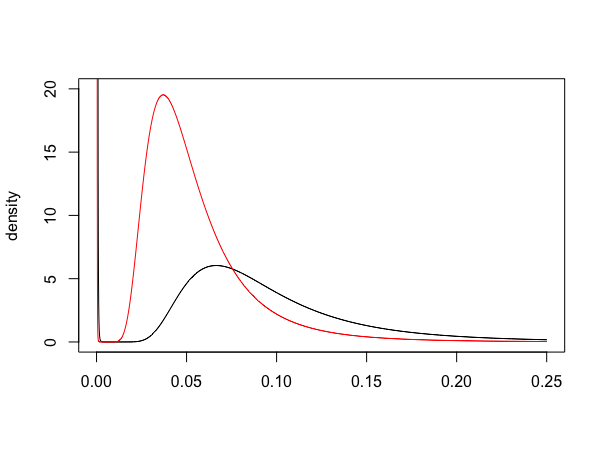}
\caption{Mixture prior for $\psi_{j,t}$ assuming the NMIG structure from \eqref{marg_beta_NMIG} with the following hyperparameters $\nu=5$, $r=0.0025$, $c_\psi=2$, $C_\psi=0.05$, $\omega_{j,1,i}=0.5$ (black) and $\omega_{j,1,i}=0.9$ (red).} \label{fig:ilustra_prioris}
\end{figure}

\noindent $j=1,\ldots,q$, where $f^{*}(w)=c[(1-\omega_{j,1,i})r+\omega_{j,1,i}]$ depends on the value of on the distribution constant $c$ from Table~\ref{tab:tab_mixture_priors} specified for $\tau_j^2|\matr{\theta}$ and on the value of $\omega_{j,1,i}=p\left(K_{j,t}=\upsilon_1|K_{j,t-1}=\upsilon_i\right)$. By defining the hyperparameters $c_\psi$ and $C_\psi$ appropriately we can learn about $Q_j$ and therefore about $\tau_j^2$. For instance, if we assume that $\nu=5$, $r=0.0025$, $c_\psi=2$,$C_\psi=0.05$, we have the following mixture densities for $\psi_{j,t}$ in Figure~\ref{fig:ilustra_prioris}, considering two values of $\omega_{j,1,i}=0.5, \, 0.9$ and the NMIG structure.

The assumption that $Q_j$ is given a prior as defined in \eqref{priorvar_dyn} makes $\tau_j^2$ also indirectly depend on the previous value of the Markov latent variable $K_{j,t-1}=\upsilon_i \in \{r,1\}$. The purpose of this is to keep the variance of the coefficients $\beta_{j,t}$ at each time point $t$ constant across the priors defined by equations \eqref{marg_beta_NMIG}, \eqref{marg_beta_NG} and \eqref{marg_beta_Laplace} and thus, comparable. 

For a simpler specification, from now on we assume that the observation variance is constant over time such that $\sigma_t^2=\sigma^2$. Extending the model to accommodate stochastic volatility is straightforward. In order to complete the specification, we shall assign prior distributions to parameters $\sigma^2$,  $\matr{\phi}$ and to the transition probabilities $\matr{\mathcal{P}}$ in a fully Bayesian strategy. For the observation variance $\sigma^2$, we assume the conjugate traditional prior
\begin{equation*}
\sigma^2 \sim \mathcal{IG}(a_\sigma,b_\sigma).
\end{equation*}  

For the AR parameters $\matr{\phi}$, we assume that each $\phi_j$ are independent from each other and distributed as 
\begin{equation*}
\phi_j \sim \mathcal{B}(a_\phi,b_\phi),
\end{equation*}  

\noindent for $j=1,\ldots,q$, where we are not considering the case $-1 < \phi_j < 0$. Finally, for the transition probabilities $\mathcal{P}_j$ we assign independent Beta distributions as
\begin{equation*}
\omega_{j,i,i} \sim \mathcal{B}(a_\omega,b_\omega),
\end{equation*}   

\noindent for $j=1,\ldots,q$, with $i \in \{0,1\}$ denoting the spike ($i=0$) or the slab ($i=1$) and $\omega_{j,k,i}=(1-\omega_{j,i,i})$, $k \neq i$, $k,i \in \{0,1\}$.

\begin{figure}[ht!]
\centering
\begin{tikzpicture}
    \node[shape=rectangle,draw=black] (A) at (0,-4) {$a_\phi,b_\phi$};
    \node[shape=rectangle,draw=black] (B) at (0,-2) {$\nu,r,a_\tau$};qaq
    \node[shape=rectangle,draw=black] (C) at (6,0) {$a_\omega,b_\omega$};
    \node[shape=rectangle,draw=black] (D) at (10,-2) {$r$};
    \node[shape=circle,draw=black] (QQ) at (2,-3) {$Q_j$};
    \node[shape=rectangle,draw=black] (QQQ) at (2,-1) {$c_\psi,C_\psi$};
    
    \node[shape=circle,draw=black] (E) at (0,-6) {$\phi_j$};
    \node[shape=circle,draw=black] (F) at (2,-5) {$\tau_j$}; 
    \node[shape=circle,draw=black] (G) at (5,-1.5) {$\omega_{j,1,1}$}; 
    \node[shape=circle,draw=black] (H) at (7,-1.5) {$\omega_{j,0,0}$};
    
    \node[shape=circle,draw=black] (I) at (4,-4) {$K_{j,2}$};
    \node[shape=circle,draw=black] (J) at (6,-4) {$K_{j,3}$}; 
    \node[shape=circle,draw=black] (S) at (8,-4) {$K_{j,4}$};
    \node[shape=circle,draw=white] (X) at (10,-4) {$\ldots$}; 
     \node[shape=circle,draw=black] (Z) at (12,-4) {$K_{j,T}$};  
    
    \node[shape=circle,draw=black] (K) at (4,-6) {$\psi_{j,2}$};
    \node[shape=circle,draw=black] (L) at (6,-6) {$\psi_{j,3}$}; 
    \node[shape=circle,draw=black] (T) at (8,-6) {$\psi_{j,4}$}; 
    \node[shape=circle,draw=white] (W) at (10,-6) {$\ldots$}; 
    \node[shape=circle,draw=black] (Y) at (12,-6) {$\psi_{j,T}$}; 
    
    \node[shape=circle,draw=black] (M) at (4,-8) {$\tilde{\beta}_{j,2}$};
    \node[shape=circle,draw=black] (N) at (6,-8) {$\tilde{\beta}_{j,3}$}; 
    \node[shape=circle,draw=black] (Q) at (2,-9) {$\sigma^2$}; 
    \node[shape=rectangle,draw=black] (R) at (0,-9) {$a_\sigma,b_\sigma$}; 
    \node[shape=circle,draw=black] (U) at (8,-8) {$\tilde{\beta}_{j,4}$}; 
     \node[shape=circle,draw=white] (YY) at (10,-8) {$\ldots$}; 
     \node[shape=circle,draw=black] (WW) at (12,-8) {$\tilde{\beta}_{j,T}$}; 
    
    \node[shape=circle,draw=black] (O) at (4,-10) {$y_{2}$};
    \node[shape=circle,draw=black] (P) at (6,-10) {$y_{3}$}; 
    \node[shape=circle,draw=black] (V) at (8,-10) {$y_{4}$}; 
    \node[shape=circle,draw=white] (KK) at (10,-10) {$\ldots$}; 
    \node[shape=circle,draw=black] (LL) at (12,-10) {$y_{T}$}; 
    
    \path [->] (A) edge node[left] {} (E);
    \path [->](B) edge node[left] {} (F);
    \path [->](C) edge node[left] {} (G);
    \path [->](C) edge node[left] {} (H);
    \path [->](G) edge node[left] {} (I);
    \path [->](G) edge node[left] {} (J);
    \path [->](H) edge node[left] {} (I);
    \path [->](H) edge node[left] {} (J);
    \path [->](D) edge node[left] {} (I);
    \path [->](D) edge node[left] {} (J);
    \path [->](I) edge node[left] {} (K);
    \path [->](J) edge node[left] {} (L);
    \path [->](F) edge node[left] {} (K);
    \path [->](F) edge node[left] {} (L);
    \path [->](K) edge node[left] {} (M);
    \path [->](L) edge node[left] {} (N);
    \path [->](E) edge node[left] {} (M);
    \path [->](E) edge node[left] {} (N);
    \path [->](M) edge node[left] {} (O);
    \path [->](N) edge node[left] {} (P);
    \path [->](Q) edge node[left] {} (O);
    \path [->](Q) edge node[left] {} (P);
    \path [->](R) edge node[left] {} (Q);
    \path [->](G) edge node[left] {} (S);
    \path [->](H) edge node[left] {} (S);
    \path [->](D) edge node[left] {} (S);
    \path [->](S) edge node[left] {} (T);
    \path [->](T) edge node[left] {} (U);
    \path [->](U) edge node[left] {} (V);
    \path [->](F) edge node[left] {} (T);
    \path [->](E) edge node[left] {} (U);
    \path [->](Q) edge node[left] {} (V);
    \path [->](F) edge node[left] {} (Y);
    \path [->](E) edge node[left] {} (WW);
    \path [->](Q) edge node[left] {} (LL);
    \path [->](G) edge node[left] {} (Z);
    \path [->](H) edge node[left] {} (Z);
    \path [->](M) edge node[left] {} (N);
    \path [->](N) edge node[left] {} (U);
    \path [->](I) edge node[left] {} (J);
    \path [->](J) edge node[left] {} (S);
    \path [->](QQQ) edge node[left] {} (QQ);
    \path [->](QQ) edge node[left] {} (F);
    \path [->](G) edge node[left] {} (QQ);
    \path [->](H) edge node[left] {} (QQ);

\end{tikzpicture}
\caption{Dependence structure for dynamic spike-and-slab model.} \label{fig:DAG_dyn_nmig}
\end{figure}
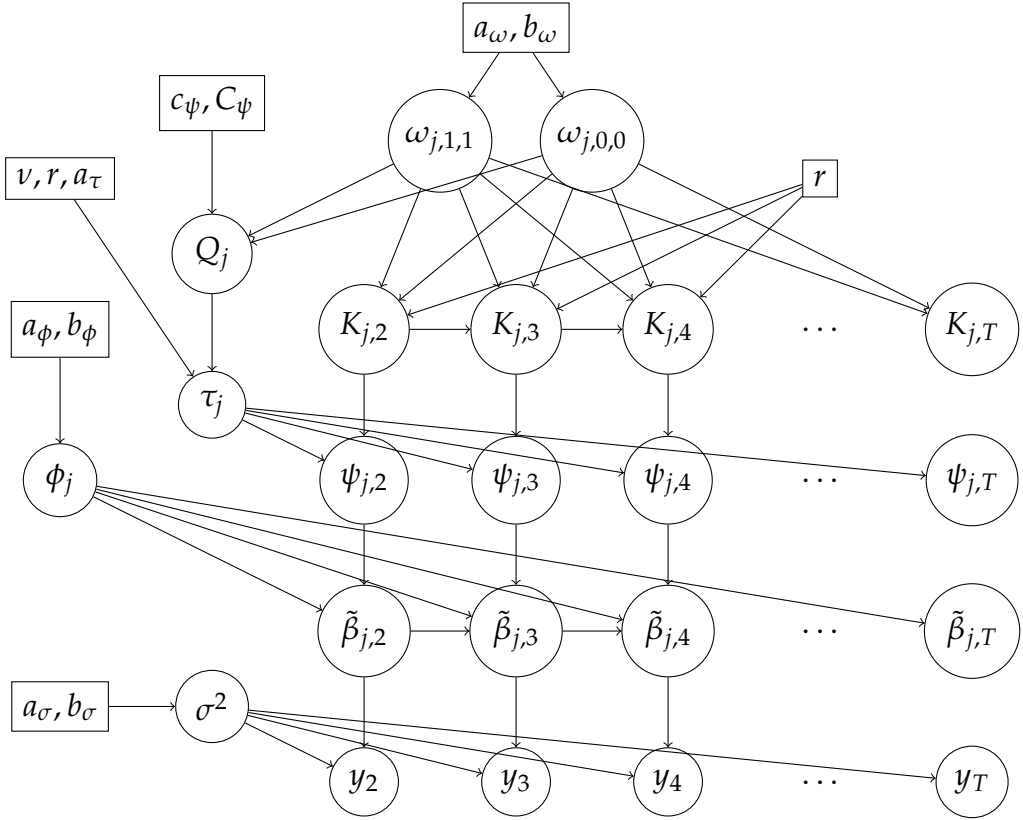

The directed acyclic graph (DAG) that summarizes the dependencies of the proposed model is shown in Figure \ref{fig:DAG_dyn_nmig}. We can see that the observations $y_1,\ldots,y_T$ are conditionally independent given the scaled coefficients $\tilde{\beta}_{j,1},\ldots,\tilde{\beta}_{j,T}$ which in turn depend on the AR parameter $\phi_j$ and on the scale $\psi_{j,t}$. The latter is driven by the Markov switching variable $K_{j,t}$, which evolves accordingly to the transition probabilities $\omega_{j,1,1}$ and $\omega_{j,0,0}$ and with $\tau_j^2$, whose distribution governs the prior choice for $\beta_{j,t}$ (e.g., mixture of Laplaces, Normal-Gammas or scaled-t).

\subsection{Posterior inference} 
\label{subsec:dynamic_spike_posterior}

The posterior distribution of the parameters can be drawn using an hybrid Gibbs sampler with an additional Metropolis-Hastings update. The scaled states $\tilde{\matr{\beta}}_1,\ldots,\tilde{\matr{\beta}}_T$ can be updated using the FFBS algorithm (due to \cite{carter1994gibbs} and \cite{fruhwirth1994data}) within the Gibbs sampler, while the process $\matr{K}=(\matr{K}_1,\ldots,\matr{K}_T)$ is updated using the algorithm proposed by \cite{gerlach2000efficient}. At each snapshot of time $t$, the probability of $K_t=1$ depends only on the previous value observed $K_{t-1}=\upsilon_{i} \in \{r,1\}$. Thus, the sequence $(K_{1},\ldots,K_{T})$ is a sequence of random variables that are Markov as for $t=2,\ldots,T$ we have $p(K_{t}|K_{1:t-1})=p(K_{t}|K_{t-1})$. 

This feature was discussed by \cite{gerlach2000efficient} in their work about \textit{dynamic mixture models}. These models adds to dynamic linear model equations \ref{dynreg} the assumption that the system matrices $\matr{F}_t$, $\matr{G}_t$ and $\matr{W}_t$ and the variance $\sigma^2_t$ are determined, up to a set of unknown parameters, by the value of $\matr{K}_t$. The algorithm and the Lemmas associated are detailed in the Appendix.

In summary, the MCMC scheme is given below.

\begin{enumerate}
\item Draw $\tilde{\matr{\beta}}$ jointly using {Forward Filtering Backward Sampling} (FFBS).

\item Draw $\matr{K}$ jointly using the algorithm of \cite{gerlach2000efficient}.

\item Draw $\sigma^2$ by its full conditional
$$
(\sigma^2|\matr{\Theta}_{\setminus \sigma^2},\matr{y}) \sim \mathcal{IG}\left(a_\sigma+\frac{T}{2},b_\sigma+\frac{1}{2}\sum_{t=1}^{T}(y_t-\matr{X}_t\matr{\beta}_{j,t})^2\right),
$$

\noindent where $\matr{\Theta}_{\setminus \sigma^2}$ denotes all the parameters to be sampled except from $\sigma^2$.

\item Draw each $\tau_j^2$ by its full conditional. Assuming the NMIG prior structure from \eqref{marg_beta_NMIG},
$$
(\tau_j^2|\matr{\Theta}_{\setminus \tau},\matr{y}) \sim \mathcal{IG}\left(\nu+\frac{T}{2},Q_j+\frac{1}{2}\sum_{t=1}^{T} \frac{\left(\beta_{j,t}-\sqrt{\dfrac{K_{j,t}}{K_{j,t-1}}}\phi_j\beta_{j,t-1}\right)^2}{K_{j,t}(1-\phi_j^2)}\right).
$$

Assuming the mixture of Laplaces from \eqref{marg_beta_NG} or the mixture of Normal-Gammas from \eqref{marg_beta_Laplace}, 

$$
(\tau_j^2|\matr{\Theta}_{\setminus \tau_j^2},\matr{y}) \sim \mathcal{GIG}(p,g,h),
$$

\noindent where 

\begin{equation*}
g=1/Q_j,\quad h=\sum_{t=1}^{T} \frac{\left(\beta_{j,t}-\sqrt{\dfrac{K_{j,t}}{K_{j,t-1}}}\phi_j\beta_{j,t-1}\right)^2}{K_{j,t}(1-\phi_j^2)}, \quad p=a_\tau-T/2.
\end{equation*}

\item Draw each $\phi_j$ using Metropolis Hastings algorithm since the full conditional 

\begin{equation*}
\begin{aligned}
p(\phi_j|\matr{\Theta}_{\setminus \phi_j},\matr{y}) &\propto p(\phi_j|a_\phi,b_\phi)p(\matr{\beta}_j|\matr{K}_j,\sigma^2,\tau_j^2)\\
&\propto\phi_j^{(a_\phi-1)}(1-\phi_j)^{(b_\phi-1)} \exp\left\{-\sum_{t=1}^{T} \frac{\left(\beta_{j,t}-\sqrt{\dfrac{\psi_{j,t}}{\psi_{j,t-1}}}\phi_j\beta_{j,t-1}\right)^2}{2\psi_{j,t}(1-\phi_j^2)}  \right\},
\end{aligned}
\end{equation*}

\noindent has no close form. We use a Beta proposal density $q\left(\phi_j^\ast|\phi_j^{(m-1)}\right)$ as

\begin{equation*}\label{phi_proposal}
\phi_j^\ast \sim \mathcal{B}\left(\alpha,\xi\left(\phi_j^{(m-1)}\right)\right), \quad \xi\left(\phi_j^{(m-1)}\right)=\alpha\left(\frac{1-\phi_j^{(m-1)}}{\phi_j^{(m-1)}}\right),
\end{equation*}

\noindent where $\alpha$ is a tuning parameter and the acceptance distribution is

$$\mathcal{A}\left(\phi_j^\ast|\phi^{(m-1)}\right)=min\left\{1,\frac{f\left(\phi_j^\ast \right)q\left(\phi_j^{(m-1)}|\phi_j^\ast\right)}{f\left(\phi_j^{(m-1)}\right)q\left(\phi_j^\ast|\phi^{(m-1)}\right)}\right\}.$$

\item Update the transition probabilities from the latent Markov process by their full conditionals

\begin{equation*}
\begin{aligned}
(\omega_{1,1}|\matr{\Theta}_{\setminus \omega_{1,1}},\matr{y}) \sim \mathcal{B}(a_\omega+\#\{t: \upsilon_1 \rightarrow \upsilon_1\},b_\omega+\#\{t: \upsilon_1 \rightarrow \upsilon_0\}),\\
(\omega_{0,0}|\matr{\Theta}_{\setminus \omega_{0,0}},\matr{y}) \sim \mathcal{B}(a_\omega+\#\{t: \upsilon_0 \rightarrow \upsilon_0\},b_\omega+\#\{t: \upsilon_0 \rightarrow \upsilon_1\}),
\end{aligned}
\end{equation*}

\noindent with $\upsilon_0=r$ and $\upsilon_1=1$.

\item Draw each $Q_j$ by its full conditional, which depends on the mixing distribution assumed for the spike-and-slab process as follows:

\begin{itemize}
\item NMIG prior: $\mathcal{GIG}(p,g,h)$, with $p=\nu-c_\psi$, $g=2\tau_j^{-2}$ and $h=2[C_\psi/f^{*}(w)],$
\item Mixture of Normal-Gammas: $\mathcal{IG}(c_\psi+a_\tau,\tau_j^{2}/2+[C_\psi/f^{*}(w)]),$
\item Mixture of Laplaces: $\mathcal{IG}(c_\psi+a_\tau,\tau_j^{2}/2+[C_\psi/f^{*}(w)])$, with $a_\tau=1$.
\end{itemize}

\end{enumerate}

\section{Synthetic and real data analyses} 
\label{simulations}

In this section we present two simulated examples where some coefficients are relevant in some periods of time and negligible in others. The first example is a singular equation model with five coefficients with four possible patterns and the second example is an application of the modified Cholesky decomposition where we simulate time-varying coefficients that compose the Cholesky factor and then apply the spike-and-slab priors on each recursive regression.  In our empirical application, we use the inflation data obtained from  Griffin's research page\footnote{Available in \url{https://www.kent.ac.uk/smsas/personal/jeg28/index.htm}}.  Inflation forecasting is a frequent topic within the shrinkage in time varying parameter models literature and was also the main subject of \cite{belmonte2014hierarchical}.

\subsection{First simulation example}

We generated the data using Equation \eqref{dyn_nmig0} with $q=5$ predictors, $T=200$ and constant observational variance $\sigma_t^2=\sigma^2=1$, where $\matr{X}_t \sim N(\matr{0},\matr{I})$ and $X_{j,1},\ldots,X_{j,T}$ are independent. We simulate the five regression coefficients as follows.

\begin{enumerate}

\item The first coefficient $\beta_{1,t}$ follows a stationary AR(1) process with AR parameter 0.97 and a Normal stationary distribution with mean 2 and variance 0.25. The initial value was drawn from its stationarity distribution $\beta_{1,1} \sim \mathcal{N}(2,0.25)$.

\item The second coefficient $\beta_{2,t}$ also follows an AR(1) process with autocorrelation parameter 0.97 and a Normal marginal distribution with mean 0 and variance 0.25, but only until the half of the sample, that is:

$$\beta_{2,t}=\left\{
                \begin{array}{ll}
                 0.97\beta_{2,t-1}+\epsilon_{2,t}, \quad &t \leq 100\\
                 0, \quad &t >100,
                \end{array}
              \right.$$ 

\noindent with the initial value drawn as $\beta_{2,1} \sim \mathcal{N}(2,0.25)$.

\item The third coefficient is always zero, except from two short periods when it equals -2:

$$\beta_{3,t}=\left\{
                \begin{array}{ll}
                 0, \quad t \leq 20; 51 \leq t \leq 120; 151\leq t\leq 200\\
                 -2, \quad 21 \leq t \leq 50; 121\leq t \leq 150.
                \end{array}
              \right.$$ 

\item The fourth coefficient is $\beta_{4,t}=0, \forall t$.

\item The fifth coefficient $\beta_{5,t}=0, \forall t$.

\end{enumerate}

\begin{figure}[ht!] 
\centering
\includegraphics[width=1.0\linewidth]{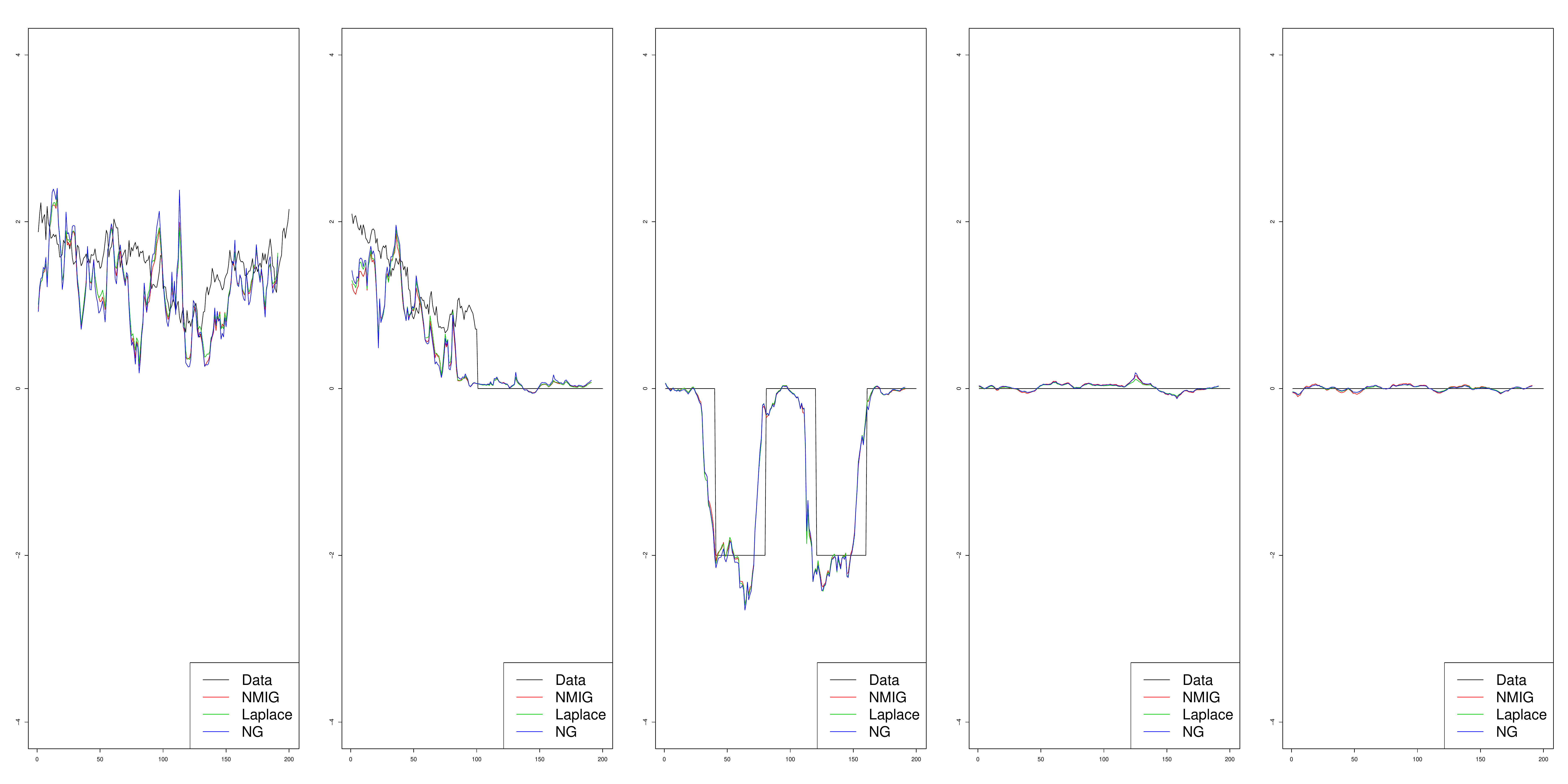} 
 \caption{Fitting of the models using posterior medians of the sampled coefficients.} 
 \label{fig:fit} 
 \end{figure}

We generate 5 replications of the data and then sample from the posterior distribution using the three mentioned priors for $\beta_{j,t}$ with the the following hyperparameters settings: $\upsilon_0=r=0.005, \upsilon_1=1, a_\tau=0.5$ (for the NG prior), $\nu=5,c_0=51,C_0=5, a_\sigma=0.0001,b_\sigma=0.0001$ (improper prior) and $\alpha=1000$ (tuning parameter for Metropolis). The MCMC algorithm was run for 10,000 iterations with half discarded as a burn-in. The prior for autoregressive parameter is $\phi_j \sim \mathcal{B}(77.6,2.4)$ for $j=1,\ldots,5$, so that it has mean 0.97.  The same choice was made for the transition probabilities $\omega_{j,0,0}$ and $\omega_{j,1,1}$. The informative prior choice is to assure that $\beta_{j,t}|\beta_{j,t-1}$ and $\psi_{j,t}|\psi_{j,t-1}$ evolves smoothly and $\matr{K}_t$ does not switch regimes so rapidly. 
Figure~\ref{fig:fit} shows the posterior medians of the coefficients $\beta_{j,t}$, comparing them to the real values, while Table~\ref{tab:RMSE_ex1} shows the root mean square error of the three priors considering the 5 replications. Figure~ref{fig:fit2} shows the posterior densities of the sampled coefficients with the Laplace prior in time points $t=10$ and $t=40$. We can note the change between these two time points: the third coefficient is equal -2 in $t=40$, so that its posterior densities shows more mass near this value. 

\begin{table}[h!]
\centering
\begin{tabular}{@{}ccc@{}}
\toprule
   & RMSE & RMSE  \\ 
Prior & (mean) & (median) \\ \midrule
NMIG    & 0.3522   & 0.3543    \\
NG      & 0.3636 & 0.3641     \\
Laplace & 0.3425  & 0.3441   \\
\bottomrule
\end{tabular}
\caption{Mean of the RMSEs of the five replications for the dynamic spike-and-slab priors using the mean and the median of the sampled coefficients - simulated example 1.}
\label{tab:RMSE_ex1}
\end{table}

\begin{figure}[ht!] 
\centering
\includegraphics[width=1.0\linewidth]{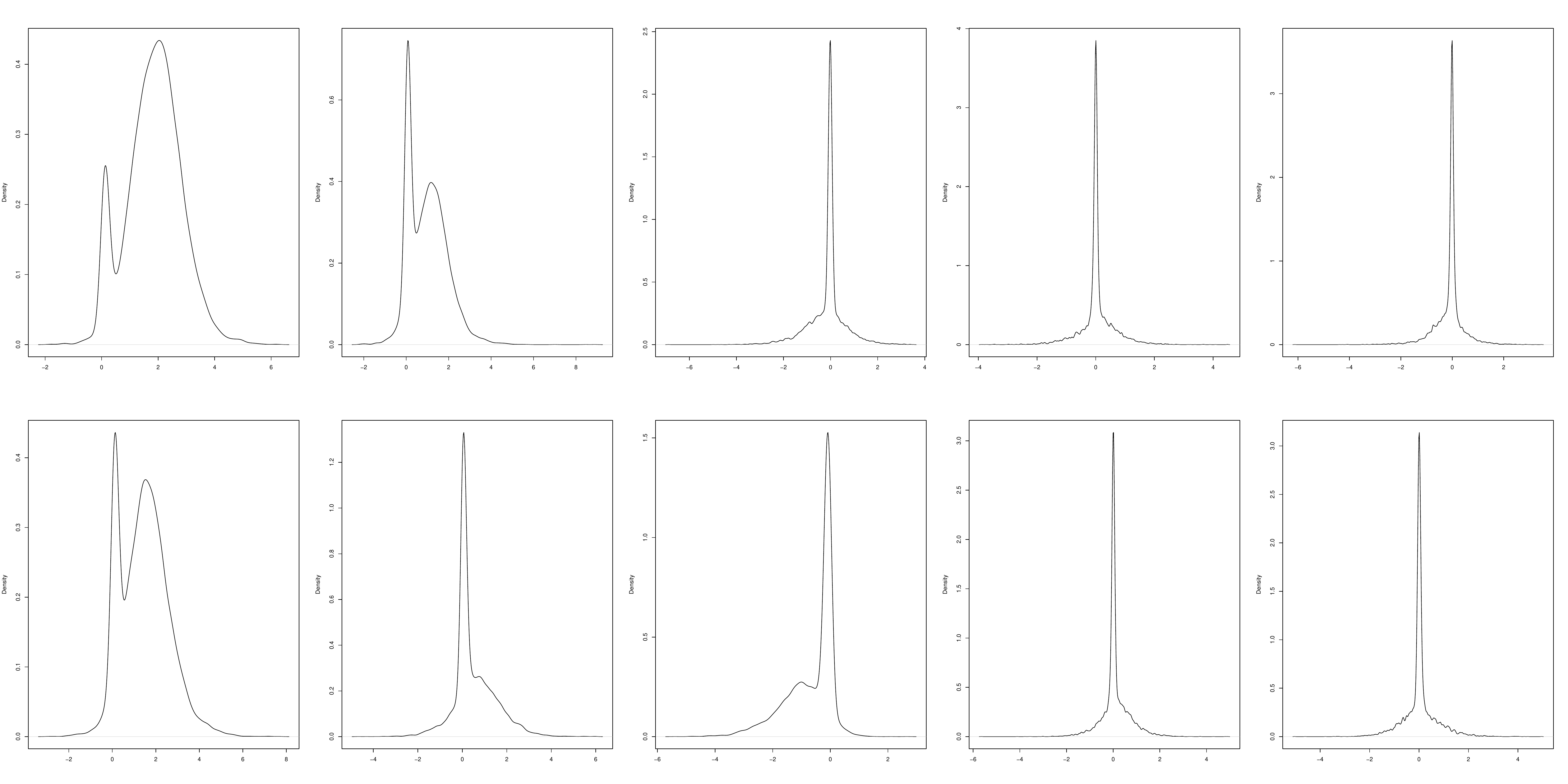}
 \caption{Posterior densities of the sampled coefficients using the dynamic Laplace prior in $t=10$ (first line) and $t=40$ (second line).} 
 \label{fig:fit2} 
 \end{figure} 

We note that the dynamic Laplace prior was slightly superior in terms of RMSE than the other two priors, but the difference is tiny. The dynamic NG with $a_\tau=0.5$ have some issues: they are much more volatile than the NMIG prior and the Laplace prior. 

\subsection{Second simulation example}

In the Cholesky decomposition each variable is regressed on its predecessors in a dynamic regression problem, that is,

$$y_{i,t}=\sum_{j=1}^{i-1}\beta_{i,j,t}y_{j,t}+\varepsilon_{i,t}$$

\noindent for $i=2,..,q$, with $y_{1,t}=\varepsilon_{1,t}$. The Cholesky factor is then 

$$\matr{T}_t=(\matr{I}-\matr{B}_t),$$

\noindent where $\matr{B}_t$ is the lower triangular matrix of coefficients for each time $t$ with zeros in the diagonal, that is, the matrix with entries $\beta_{2,1,t},\beta_{3,1,t},\beta_{3,2,t},\ldots,\beta_{q,1,t},\ldots,\beta_{q,q-1,t}$. Thus, we have $q(q-1)T$ parameters to be estimated. 

In this second example we simulate time-varying coefficients that compose the Cholesky factor $\matr{T}_t$ and then apply the spike-and-slab priors on each recursive regression. The simulation is done as follows. We define that the number of time points $T=240$ and the number of ordered variables that compose the vector $\matr{y}$ is $q=10$. We sample from four possible processes for the time-varying coefficients with the same probability of occurrence. They are:

\begin{enumerate}

\item A stationary AR(1) process with autoregressive coefficient $\phi=0.98$ and with fixed variance $\sigma^2=(1-\phi)0.15$, without an intercept term, that is

$$\beta_{i,j,t}=\phi \beta_{i,j,t-1}+\nu_{i,j,t},$$ 

\noindent with $\nu_{i,j,t} \sim \mathcal{N}(0,\sigma^2)$. 

\item A stationary AR(1) process with autoregressive coefficient $\phi=0.98$ and with fixed variance $\sigma^2=(1-\phi)0.15$ until the half of the time points. Then, the coefficient is set to zero.

\item A fixed interval process similar to the third coefficient from the first simulated example as follows

$$\beta_{i,j,t}=\left\{
                \begin{array}{cc}
                 0, \quad t \leq T/8; 3T/8 < t \leq 5T/8; t >7T/8 \\
                 -0.5, \quad T/8 \leq t < 3T/8; 5T/8 < t \leq 7T/8.
                \end{array}
              \right.$$ 

\item A constant coefficient equal to zero.

\end{enumerate}

In this manner, we want to give a structure to the Cholesky factor, but now allowing for time-varying coefficients. Each coefficient $\beta_{i,j,t}$ follows one of the four processes: (1) AR(1), (2) AR(1) with zeros, (3) fixed intervals, or (4) zeros, which are sampled using equal probabilities. Then, we build the 10 time series $\matr{y}_1,\ldots,\matr{y}_{10}$ as

\begin{equation*}
\begin{aligned}
&y_{1,t}=\varepsilon_{1,t}\\
&y_{2,t}=\beta_{2,1,t}y_{1,t}+\varepsilon_{2,t}\\
& \ldots\\
&y_{10,t}=\sum_{j=1}^{9}\beta_{10,j,t}y_{j,t}+\varepsilon_{10,t},\\
\end{aligned}
\end{equation*}

\noindent for $t=1,..,240$ and where $\varepsilon_{i,t} \sim \mathcal{N}(0,0.0625)$, $\forall j=1,..,10$.

The results for the RMSE are shown in Table \ref{tab:RMSE_ex2}. The MCMC scheme uses 10,000 simulations with $5,000$ discarded as burn-in. The hyperparameters were set as follows: $\upsilon_0=r=0.005, \upsilon_1=1, a_\tau=0.5$ (for the NG prior), $\nu=25,c_0=50,C_0=1.5, a_\sigma=5,b_\sigma=1.5$ and $\alpha=1000$ (tuning parameter for Metropolis). 

\begin{table}[ht!]
\centering
\begin{tabular}{@{}ccc@{}}
\toprule
   & RMSE & RMSE  \\ 
Prior & (mean) & (median) \\ \midrule
NMIG    & 0.2472   & 0.2863    \\
NG      & 0.2398  & 0.2820    \\
Laplace & 0.2401 & 0.2842   \\ 
\bottomrule
\end{tabular}
\caption{RMSE for the dynamic spike-and-slab priors using the mean and the median of the sampled coefficients - simulated example 2}
\label{tab:RMSE_ex2}
\end{table}

\subsection{Predicting inflation} 
\label{applications}

The empirical application uses inflation data obtained from Professor Griffin's research page\footnote{Available in \url{https://www.kent.ac.uk/smsas/personal/jeg28/index.htm}}. We use the inflation data collected by them with the independent variable as the US quarterly inflation measure based on the Gross Domestic Product (GDP). The data was obtained from FRED database, Federal Reserve Bank of St.Louis, University of Michigan Consumer Survey database, Federal Reserve Bank of Philadelphia, and Institute of Supply Management. The data set includes 31 predictors, from activity and term structure variables to survey forecasts and previous lags. A full description of the 31 explanatory variables can be found in Appendix. The sample period is from the second quarter of 1965 to first quarter of 2011 with $T=182$ observations.

Inflation forecasting is a frequent topic within the shrinkage in time varying parameter models literature and was also the main subject of \cite{belmonte2014hierarchical}. The size of the set of potential variables to forecast inflation is huge and, as noted by \cite{kalli2014time}, this is usually split into four subsets: past inflation forecasts, where the explanatory variables are previous lags of inflation; Phillips curve forecasts, which involve activity variables, such as economic growth rate or output gap, unemployment rate, and lagged inflation; forecasts based on variables which are themselves forecasts of asset prices (combination indices), term structures of nominal debt, and consumer surveys; and forecasts based on other exogenous variables such as government investment, the number of new private houses.

We applied the three variable selection priors  \eqref{marg_beta_NMIG}, \eqref{marg_beta_NG} and \eqref{marg_beta_Laplace} to the GDP deflator data with the following hyperparameter settings: $\upsilon_0=r=0.05, \upsilon_1=1, a_\tau=0.5$ (for the NG prior), $\nu=50,c_0=50,C_0=.05, a_\sigma=31,b_\sigma=30\hat{\sigma}^2=4.22$, with $\hat{\sigma}^2=0.14$ being the sum of square residuals of the OLS estimate divided by $(T-1)$ and $\alpha=1000$ (tuning parameter for Metropolis). The previous Beta priors, that is, $\phi_j \sim \mathcal{B}(77.6,2.4)$ and for the transition probabilities $\omega_{j,0,0} \sim \mathcal{B}(77.6,2.4)$ and $\omega_{j,1,1} \sim \mathcal{B}(77.6,2.4)$ were maintained. We ran a total of 20,000 iterations of the MCMC scheme and we discarded 10,000 as a burn-in. 

The results (the mean of the coefficients $\beta_{j,t}$) were compared to results from the NGAR process defined \cite{kalli2014time}, see Figure \ref{fig:compare_NMIG2} below. We used the MATLAB code provided by Professor Griffin in his website for the GDP inflation data after standardizing both the response and the predictors in the same way as done by the authors. Figure \ref{fig:compare_NMIG2} indicates, the results are quite similar.  Despite its simplicity our proposed method, which is more computationally fast, produces competitive results.

\begin{figure}[H]
\centering
\includegraphics[width=1.0\linewidth]{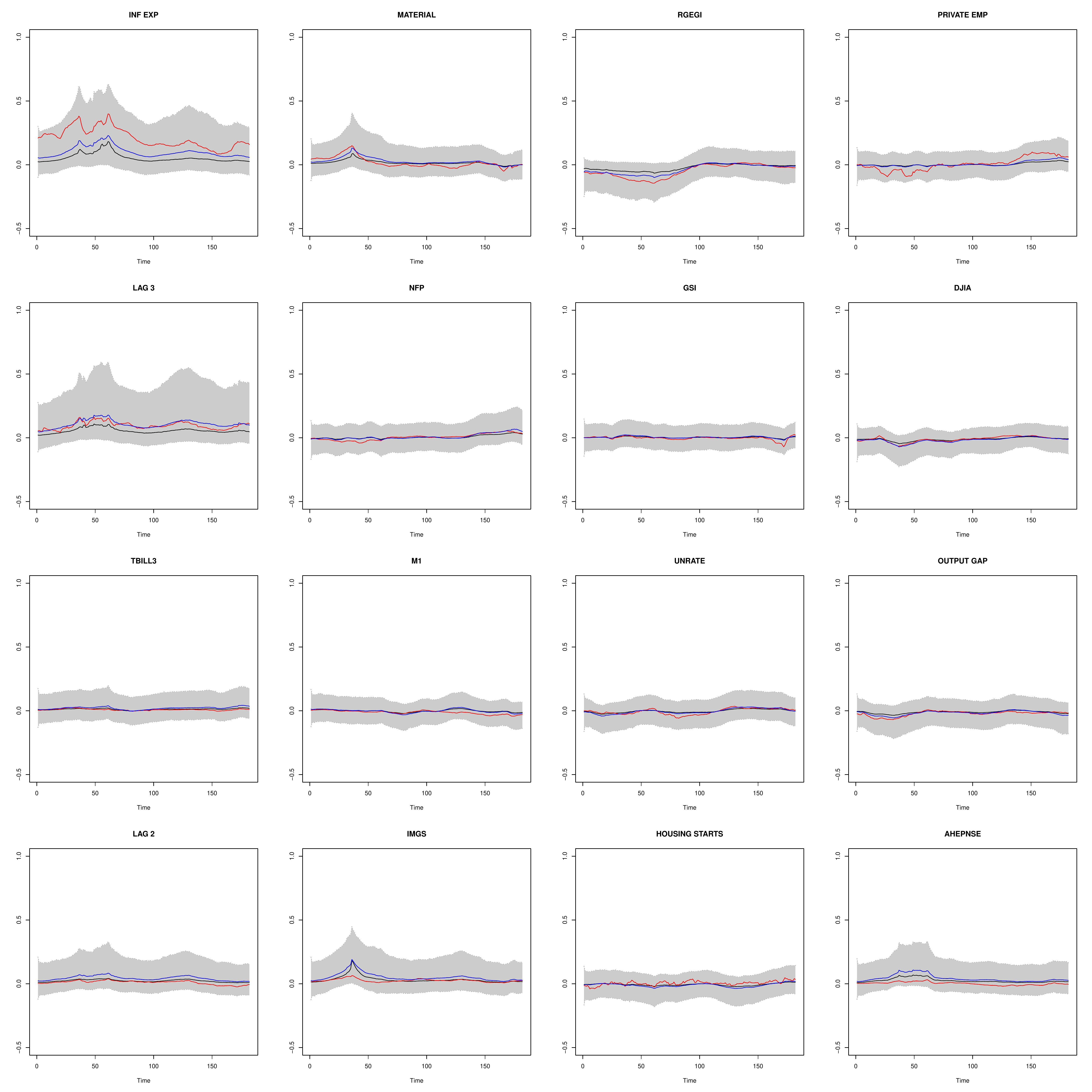} 
 \caption{Comparison between the mean of the NGAR model and the dynamic NMIG prior relevances. Mean NGAR=red line; Mean NMIG=blue line; Median NMIG=black line; 95\% confidence intervals=grey area.} 
 \label{fig:compare_NMIG2} 
 \end{figure}

 \newpage
 
 \section{Conclusions}
 \label{conclusion}

We introduce a novel strategy to allow dynamic sparsity, or vertical sparsity, in dynamic regression models, with particular interest in the time-varying Cholesky decomposition set up by \cite{lopes2008}.
Our scheme allows for time-varying sparsity, based on an extension of spike-and-slab priors for dynamic models using Markov switching auxiliary variables. 
It is simpler than the existing ones, such as \cite{nakajima2013bayesian} and \cite{kalli2014time}, while maintaining scalability.  
It also leads to a more efficient MCMC as time-varying latent variables $\tilde{\matr{\beta}}=(\tilde{\matr{\beta}}_1,\ldots,\tilde{\matr{\beta}}_T)$ and $\matr{K}=(\matr{K}_1,\ldots,\matr{K}_T)$ 
can be sampled jointly by combining a marginal Gibbs step for $\matr{K}$ (\cite{gerlach2000efficient}) with a conditional FFBS step for $\tilde{\matr{\beta}}|\matr{K}$ (\cite{fruhwirth1994data}).  
The alternative ones sample from full conditional distributions, which is notoriously known to lead to slow mixing in dynamic systems.

\bibliography{uribe-lopes-references}
\bibliographystyle{apalike}

\newpage

\section*{Appendix I: Efficient Bayesian inference for dynamic mixtures} 
\label{appendixI}

We present here in details the dynamic mixture model approach proposed by \cite{gerlach2000efficient}. Consider the univariate Gaussian DLM as

\begin{equation}\label{gck_dlm}
\begin{aligned}
y_t&=f_t+\matr{F}_t'\matr{\theta}_t+\gamma_t u_t, \qquad u_t \sim \mathcal{N}(0,1)  \\
\matr{\theta}_t&=\matr{g}_t+\matr{G}_t\matr{\theta}_{t-1}+\matr{\Gamma}\matr{v}_t,\qquad \matr{v}_t \sim \mathcal{N}(\matr{0},\matr{I}),
\end{aligned}
\end{equation}

\noindent for $t=1,..,n$, where $\matr{\theta}_t$ is a $q$-dimensional vector of states, $u_t$ and $\matr{v}_t$ are independent and standard Normal distributed, and $f_t$, $\matr{F}_t'$, $\gamma_t$, $\matr{g}_t$, $\matr{G}_t$ and $\matr{\Gamma}_t$ may all depend on the vector Markov $\matr{K}_t$ and on a vector of parameters $\matr{\Phi}$. This makes observations $y_t$ mixture of normals.

Note that $\matr{K}_{1:n}=(\matr{K}_1,\ldots,\matr{K}_n)$ is a sequence of random vectors that are Markov:
$$
p(\matr{K}_t|\matr{K}_{1:t-1})=p(\matr{K}_t|\matr{K}_{t-1}), \qquad t=2,\ldots,n.
$$

The sampling scheme proposed generates $\matr{K}_t$ from density $p(\matr{K}_t|y_{1:n},\matr{K}_{s \neq t})$ for $t=1,\ldots,n$ without conditioning on the states $\matr{\theta}_{1:n}$. The crucial thing is to notice that

\begin{equation}\label{gck1}
\begin{aligned}
p(\matr{K}_t|y_{1:n},\matr{K}_{s \neq t} &\propto p(y_{1:n}|\matr{K}_{1:n})p(\matr{K}_t|\matr{K}_{s \neq t})\\
&\propto p(y_{t+1:n}|y_{1:t},\matr{K}_{1:n})p(y_t|y_{1:t-1},\matr{K}_{1:t})p(\matr{K}_t|\matr{K}_{s \neq t}),
\end{aligned}
\end{equation}

\noindent where the dependence on the parameters $\matr{\Phi}$ has been suppressed for convenience. 

For each value of $\matr{K}_t$ the right size of \eqref{gck1} is evaluated as follows. The term $p(\matr{K}_t|\matr{K}_{s \neq t})$ is obtained from the prior. The term $p(y_t|y_{1:t-1},\matr{K}_{1:t})$ is obtained from $p(\matr{\theta}_{t-1}|y_{1:t-1},\matr{K}_{1:t-1})$, i.e., from the the filtering distribution, using one step of the Kalman filter.

Obtaining the term $p(y_{t+1:n}|y_{1:t},\matr{K}_{1:n})$ is the crucial innovation of the algorithm of \cite{gerlach2000efficient}. Traditional sampling algorithms use $n-t +1$ steps of the Kalman filter given the current values of $\matr{K}_{t,n}$ to obtain the term $p(y_{t+1:n}|y_{1:t},\matr{K}_{1:n})$. Therefore, it requires $\mathcal{O}(n)$ operations to generate each $\matr{K}_t$, and hence $\mathcal{O}(n^2)$ operations to generate $\matr{K}_{1:n}$. Nevertheless, in the proposed algorithm the term $p(y_{t+1:n}|y_{1:t},\matr{K}_{1:n})$ is obtained in one step after an initial set of backward recursions. This reduces the number of operations required to generate the complete vector $\matr{K}_{1:n}$ to $\mathcal{O}(n)$.

Before giving the efficient method for generating $\matr{K}_{1:n}$, we are going to state several preliminary lemmas, whose proofs can be found in the Appendix of  \cite{gerlach2000efficient}. All of the lemmas refer to the univariate Gaussian DLM \eqref{gck_dlm}.

\begin{lemma}{}\label{lemma1} Let $r_{t+1}=\Var(y_{t+1}|\matr{\theta}_t,\matr{K}_{1:t+1})$. Then, the following hold:

$$\mathbb{E}(y_{t+1}|\matr{\theta}_t,\matr{K}_{1:t+1})=f_{t+1}+\matr{F}_{t+1}'(\matr{g}_{t+1}+\matr{G}_{t+1}\matr{\theta}_t),$$
 $$r_{t+1}=\matr{F}_{t+1}'\matr{\Gamma}_{t+1}\matr{\Gamma}_{t+1}'\matr{F}_{t+1}+\gamma_{t+1}^2,$$

\noindent and

$$\mathbb{E}(\matr{\theta}_{t+1}|\matr{\theta}_t,y_{t+1},\matr{K}_{1:n})=\matr{a}_{t+1}+\matr{A}_{t+1}\matr{\theta}_t+\matr{B}_{t+1}y_{t+1},$$
$$\Var(\matr{\theta}_{t+1}|\matr{\theta}_t,y_{t+1},\matr{K}_{1:n})=\matr{C}_{t+1}\matr{C}_{t+1}',$$

\noindent where 

$$\matr{a}_{t+1}=(\matr{I}-\matr{B}_{t+1}\matr{F}_{t+1}')\matr{g}_t-\matr{B}_{t+1}f_t,$$
$$\matr{A}_{t+1}=(\matr{I}-\matr{B}_{t+1}\matr{F}_{t+1}')\matr{G}_{t+1},$$
$$\matr{B}_{t+1}=\matr{\Gamma}_{t+1}\matr{\Gamma}_{t+1}'\matr{F}_{t+1}r_{t+1}^{-1},$$
$$\matr{C}_{t+1}\matr{C}_{t+1}'=\matr{\Gamma}_{t+1}(\matr{I}-\matr{\Gamma}_{t+1}'\matr{F}_{t+1}r_{t+1}^{-1}\matr{F}_{t+1}' \matr{\Gamma}_{t+1})\matr{\Gamma}_{t+1},$$

It is straightforward to factor the expression on the right side of the last equality to get a matrix $\matr{C}_{t+1}$ that either is null or has full column rank. Then, we can write

$$\matr{\theta}_{t+1}=\matr{a}_{t+1}+\matr{A}_{t+1}+\matr{B}_{t+1}y_{t+1}+\matr{C}_{t+1}\matr{\xi}_{t+1},$$

\noindent where $\matr{\xi}_{t+1} \sim \mathcal{N}(\matr{0},\matr{I})$ and is independent of $\matr{\theta}_t$ and $y_{t+1}$, conditional on $\matr{K}_{1:n}$.

\end{lemma}

\begin{lemma}{}\label{lemma2} For $t=1,\ldots,n-1$, the density $p(y_{t+1:n}|\matr{\theta}_t,\matr{K}_{1:n})$ is independent of $\matr{K}_{1:t}$ and can be expressed as
$$
p(y_{t+1:n}|\matr{\theta}_t,\matr{K}_{1:n}) \propto \exp \left \{-\frac{1}{2}(\matr{\theta}_t\matr{\Omega}_t(\matr{\theta}_t-2\matr{\mu}_t'\matr{\theta}_t)   \right\},
$$
where the terms $\matr{\Omega}_t$ and $\matr{\mu}_t$ are computed recursively starting from
$$
\matr{\Omega}_n=\matr{0}, \qquad \matr{\mu}_n=\matr{0}.
$$
Updating backward, we obtain
\begin{eqnarray*}
\matr{\Omega}_t &=& \matr{A}_{t+1}'(\matr{\Omega}_{t+1}-\matr{\Omega}_{t+1} \matr{C}_{t+1}\matr{D}_{t+1}^{-1}\matr{C}_{t+1}'\matr{\Omega}_{t+1})\matr{A}_{t+1}+\matr{G}_{t+1}'\matr{F}_{t+1}r_{t+1}^{-1}\matr{F}_{t+1}'\matr{G}_{t+1},\\
\matr{\mu}_t &=& \matr{A}_{t+1}'(\matr{I}-\matr{\Omega}_{t+1}\matr{C}_{t+1}\matr{D}_{t+1}^{-1}\matr{C}_{t+1}')(\matr{\mu}_{t+1}-\matr{\Omega}_{t+1}(\matr{a}_{t+1}+\matr{B}_{t+1}y_{t+1}))\\
&+&\matr{G}_{t+1}'\matr{F}_{t+1}r_{t+1}^{-1}(y_{t+1}-f_{t+1}-\matr{F}_{t+1}'\matr{g}_{t+1}),\\
\matr{D}_{t+1} &=& \matr{C}_{t+1}'\matr{\Omega}_{t+1}\matr{C}_{t+1}+\matr{I}.
\end{eqnarray*}
\end{lemma}

\begin{lemma}{}\label{lemma3} Let $\matr{m}_t=\mathbb{E}(\matr{\theta}_t|y_{1:t},\matr{K}_{1:n})$, $\matr{V}_t=\Var(\matr{\theta}_t|y_{1:t},\matr{K}_{1:n})$ and $R_t=\Var(y_t|y_{1:t-1},\matr{K}_{1:n})$. The Kalman filter for the model \eqref{gck_dlm} is given by
\begin{eqnarray*}
\matr{R}_t &=& \matr{F}_t'\matr{G}_t\matr{V}_{t-1}\matr{G}_t'\matr{F}_t+\matr{F}_t'\matr{\Gamma}_t\matr{\Gamma}_t'\matr{F}_t+\gamma_t^2,\\
\matr{m}_t &=& (\matr{I}-\matr{J}_t\matr{F}_t')(\matr{g}_t+\matr{G}_t\matr{m}_{t-1})+\matr{J}_t(y_t-f_t),\\
\matr{V}_t &=& \matr{G}_t\matr{V}_{t-1}\matr{G}_t'+\matr{\Gamma}_t\matr{\Gamma}_t'-\matr{J}_t\matr{J}_t'R_t,
\end{eqnarray*}
where
$$
\matr{J}_t=[\matr{G}_t\matr{V}_{t-1}\matr{G}_t'\matr{F}_t+\matr{\Gamma}_t\matr{\Gamma}_t'\matr{F}_t]/R_t.
$$
The conditional density $p(y_t|y_{1:t-1},\matr{K}_{1:t})$ is such that
$$
 p(y_t|y_{1:t-1},\matr{K}_{1:t}) \propto R_t^{-1}\exp \left\{-\frac{1}{2R_t}(y_t-f_t-\matr{F}_t'(\matr{g}_t+\matr{G}_t\matr{m}_{t-1}))^2  \right\}.
$$

We can write $\matr{V}_t=\matr{T}_t\matr{T}_t'$, where the matrix $\matr{T}_t$ either has full column rank if $\matr{V}_t \neq \matr{0}$ or is null if $\matr{V}_t=\matr{0}$. Conditional on $\matr{K}_{1:n}$, we can express $\matr{\theta}_t$ as

$$\matr{\theta}_t=\matr{m}_t+\matr{T}_t\matr{\xi}_t,$$

\noindent where $\matr{\xi}_t \sim \mathcal{N}(\matr{0},\matr{I})$ and is independent of $y_{1:t}$.

\end{lemma}

The next Lemma uses Lemma \ref{lemma3} to efficiently evaluate the factor $p(y_{t+1:n}|y_{1:t},\matr{K}_{1:n})$.\\

\begin{lemma}{}\label{lemma4} Using the results of Lemma \ref{lemma3}, it follows that

\begin{equation*}
\begin{aligned}
p(y_{t+1:n}|y_{1:t},\matr{K}_{1:n}) &= \int p(y_{t+1:n}|\matr{\theta}_t,\matr{K}_{t+1:n})p(\matr{\xi}_t|\matr{K}_{1:t})d\matr{\xi}_t\\
& \propto |\matr{T}_t'\matr{\Omega}_t\matr{T}_t+\matr{I}|^{-1/2}\exp \left \{ -\frac{1}{2}\left(\matr{m}_t'\matr{\Omega}_t\matr{m}_t-2\matr{\mu}_t'\matr{m}_t-\matr{\phi}_t'(\matr{T}_t'\matr{\Omega}_t\matr{T}_t+\matr{I})^{-1}\matr{\phi}_t\right) \right \},
\end{aligned}
\end{equation*}

\noindent where $\matr{\phi}_t=\matr{T}_t'(\matr{\mu}_t-\matr{\Omega}_t\matr{m}_t)$.

\end{lemma}

The recursion for generating $\matr{K}_{1:n}$ in $\mathcal{O}(n)$ operations is now given.

\RestyleAlgo{boxruled}
\begin{algorithm}[ht!]
  \caption{The algorithm of \cite{gerlach2000efficient} for dynamic mixture models}\label{gck_algo}

1. Given the current value of $\matr{K}_{1:n}$, calculate $\matr{\Omega}_t$ and $\matr{\mu}_t$ for $t=n-1,\ldots,1$, using the recursions in Lemma \ref{lemma2}.

2. Given $\mathbb{E}(\matr{\theta}_0)$ and $\Var(\matr{\theta}_0)$, perform the following for $t =1,\ldots  n$:

(a) Obtain $R_t$, $\matr{m}_t$ and $\matr{V}_t$ from $\matr{m}_{t-1}$ and $\matr{V}_{t-1}$ as in Lemma \ref{lemma3};

(b) Obtain $p(y_t|y_{1:t-1},\matr{K}_{1:t})$ as in Lemma  \ref{lemma3} and $p(y_{t+1:n}|y_{1:t},\matr{K}_{1:n})$ as in Lemma \ref{lemma4};

(c) Obtain $p(\matr{K}_t|y_{1:n}\matr{K}_{s \neq t})$ for all values of $\matr{K}_t$ by normalization of

\begin{equation*}
\begin{aligned}
p(\matr{K}_t|y_{1:n},\matr{K}_{s \neq t} &\propto p(y_{1:n}|\matr{K}_{1:n})p(\matr{K}_t|\matr{K}_{s \neq t})\\
&\propto p(y_{t+1:n}|y_{1:t},\matr{K}_{1:n})p(y_t|y_{1:t-1},\matr{K}_{1:t})p(\matr{K}_t|\matr{K}_{s \neq t}).
\end{aligned}
\end{equation*}

Then, draw $\matr{K}_t$.

(d) Update $\matr{m}_t$ and $\matr{V}_t$ as in Lemma \ref{lemma3}, using the generated value of  $\matr{K}_t$.

\end{algorithm}

\newpage

\section*{Appendix II: inflation data} 
\label{appendixII}

\begin{table}[ht!]
\centering
\begin{tabular}{p{5cm}p{10cm}}
\toprule
Name & Description\\ \midrule
GDP & Difference in logs of real gross domestic product\\
PCE & Difference in logs of real personal consumption expenditure\\
GPI & Difference in logs of real gross private investment\\
RGEGI & Difference in logs of real government consumption expenditure and gross investment\\
IMGS & Difference in logs of imports of goods and services\\
NFP & Difference in logs non-farm payroll\\
M2 & Difference in logs M2 (commercial bank money)\\
ENERGY & Difference in logs of oil price index\\
FOOD & Difference in logs of food price index\\
MATERIALS & Difference in logs of producer price index (PPI) industrial commodities\\
OUTPUT GAP & Difference in logs of potential GDP level\\
GS10 & Difference in logs of 10yr Treasury constant maturity rate\\
GS5 & Difference in logs of 5yr Treasury constant maturity rate\\
GS3 & Difference in logs 3yr Treasury constant maturity rate\\
GS1 & Difference in logs 1yr Treasury constant maturity rate\\
PRIVATE EMPLOYMENT & Log difference in total private employment\\
PMI MANU & Log difference in PMI-manufacturing index\\
AHEPNSE & Log difference in average hourly earnings of private non management employees\\
DJIA & Log difference in Dow Jones Industrial Average Returns\\
M1 & Log difference in M1 (narrow-commercial bank money)\\
ISM SDI & Institute for Supply Management (ISM) Supplier Deliveries Inventory\\
CONSUMER & University of Michigan consumer sentiment (level)\\
UNRATE & Log of the unemployment rate\\
TBILL3 3m & Treasury bill rate\\
TBILL SPREAD & Difference between GS10 and TBILL3\\
HOUSING STARTS & Private housing (in thousands of units)\\
INF EXP & University of Michigan inflation expectations (level)\\
LAG1, LAG2, LAG3, LAG4 & The first, second, third and fourth lag\\
\bottomrule
\end{tabular}
\caption{Inflation Data. Sources: FRED database, Federal Reserve Bank of St.Louis, University
of Michigan Consumer Survey database, Federal Reserve Bank of Philadelphia, and Institute of
Supply Management.}
\label{tab:inflation_data}
\end{table}

\end{document}